\documentclass[10pt,conference]{IEEEtran}
\IEEEoverridecommandlockouts

\usepackage{cite}
\usepackage{amsmath,amssymb,amsfonts}
\usepackage{algorithmic}
\usepackage{graphicx}
\usepackage{textcomp}
\usepackage{xcolor}

\usepackage{caption}
\usepackage{subcaption}  
\usepackage{enumitem}
\usepackage[english]{babel}
\usepackage{amsthm}

\usepackage{microtype}

\usepackage{pgfplots}
\usepackage{tikz}
\usetikzlibrary{calc}
\makeatletter
\newcommand{\gettikzxy}[3]{%
  \tikz@scan@one@point\pgfutil@firstofone#1\relax
  \edef#2{\the\pgf@x}%
  \edef#3{\the\pgf@y}%
}
\usetikzlibrary{spy,backgrounds}
\usepackage{pgfplotstable}

\pgfplotsset{compat=newest}
\usetikzlibrary{plotmarks}
\usetikzlibrary{arrows.meta}
\usepgfplotslibrary{patchplots}
\usepackage{grffile}
\newlength\fheight 
\newlength\fwidth 
\usepgfplotslibrary{fillbetween}

\usepackage{acronym}

\acrodef{ot}[OT]{optimal transport}
\acrodef{csi}[CSI]{channel state information}

\acrodef{5g}[5G]{the fifth generation}
\acrodef{6g}[6G]{the sixth generation}
\acrodef{iot}[IoT]{internet of things}

\acrodef{aoa}[AoA]{angle-of-arrival}
\acrodef{toa}[ToA]{time-of-arrival}
\acrodef{rtt}[RTT]{round-trip time}

\acrodef{bs}[BS]{base station}
\acrodef{cdf}[CDF]{cumulative distribution function}
\acrodef{crb}[CRB]{Cram\'er-Rao bound}
\acrodef{gnss}[GNSS]{global navigation satellite system}
\acrodef{los}[LoS]{line-of-sight}
\acrodef{nlos}[NLoS]{non-line-of-sight}
\acrodef{mae}[MAE]{mean absolute value}
\acrodef{music}[MUSIC]{multiple signal classification}
\acrodef{ofdm}[OFDM]{orthogonal frequency division multiplexing}
\acrodef{ue}[UE]{user equipment}
\acrodef{upa}[UPA]{uniform planar array}
\acrodef{ula}[ULA]{uniform linear array}

\acrodef{ml}[ML]{machine learning}
\acrodef{nn}[NN]{neural network}
\acrodef{cnn}[CNN]{convolutional neural network}

\acrodef{music}[MUSIC]{multiple signal classification}
\acrodef{omp}[OMP]{orthogonal matching pursuit}
\acrodef{mae}[MAE]{mean absolute error}
\acrodef{cdf}[CDF]{cumulative distribution function}

\definecolor{myColor}{RGB}{0, 100, 0}

\newtheorem*{remark}{\textbf{Remark}}

\usepackage{titlesec}
\titlespacing{\section}{-10pt}{*0}{*0} 

\usepackage{geometry}
\begin{document}


\title{UNILoc: Unified Localization Combining Model-Based Geometry and Unsupervised Learning
\thanks{This work has been supported, in part, by the European Union through the project ISLANDS - Grant agreement n. 101120544, by the Swedish Research Council (VR) through the project 6G-PERCEF under Grant 2024-04390, and by Business Finland under the 6G-ISAC project. E-mail: \{yuhaozh, guangjin.pan, furkan, henkw\}@chalmers.se; \{ossi.kaltiokallio, mikko.valkama\}@tuni.fi.}
}

\author{\IEEEauthorblockN{Yuhao Zhang\IEEEauthorrefmark{1}, Guangjin Pan\IEEEauthorrefmark{1}, Musa Furkan Keskin\IEEEauthorrefmark{1}, Ossi Kaltiokallio\IEEEauthorrefmark{2}, Mikko Valkama\IEEEauthorrefmark{2}, Henk Wymeersch\IEEEauthorrefmark{1}}
\IEEEauthorblockA{
\IEEEauthorrefmark{1}Department of Electrical Engineering, Chalmers University of Technology, Sweden \\
\IEEEauthorrefmark{2}Unit of Electrical Engineering, Tampere University, Finland
}}

\maketitle

\begin{abstract}
Accurate mobile device localization is critical for emerging 5G/6G applications such as autonomous vehicles and augmented reality. In this paper, we propose a unified localization method that integrates model-based and \ac{ml}-based methods to reap their respective advantages by exploiting available map information. In order to avoid supervised learning, we generate training labels automatically via \ac{ot} by fusing geometric estimates with building layouts. Ray-tracing based simulations are carried out to demonstrate that the proposed method significantly improves positioning accuracy for both \ac{los} users (compared to \ac{ml}-based methods) and \ac{nlos} users (compared to model-based methods). Remarkably, the unified method is able to achieve competitive overall performance with the fully-supervised fingerprinting, while eliminating the need for cumbersome labeled data measurement and collection.
\end{abstract}


\acresetall

\section{Introduction}

Accurate localization of mobile devices is critical for a wide range of applications, including emergency services, navigation, autonomous driving, and industrial \ac{iot}~\cite{A_Tuto_Chen_2022,Posi_And_Shah_2018,Harn_NLOS_Mend_2019}. While \ac{gnss} provides a viable solution for rural outdoor scenarios, their effectiveness degrades significantly in dense urban environments and indoor settings due to signal blockage and multipath interference. Wireless communication systems, e.g., 5G/6G networks, present new opportunities for high-precision positioning by leveraging, for example, wide bandwidths, potentially high carrier frequency, and directional beamforming to extract fine-grained \ac{csi}~\cite{A_Tuto_Chen_2022,Posi_And_Shah_2018,Harn_NLOS_Mend_2019,An_Intr_Zeka_2012,Robu_Snap_Kalt_2024}.

\begin{figure}[t]
    \centering
    \begin{tikzpicture}[every node/.style={font=\footnotesize}]
    \node (image) [anchor=south west]{\includegraphics[width=0.6\linewidth]{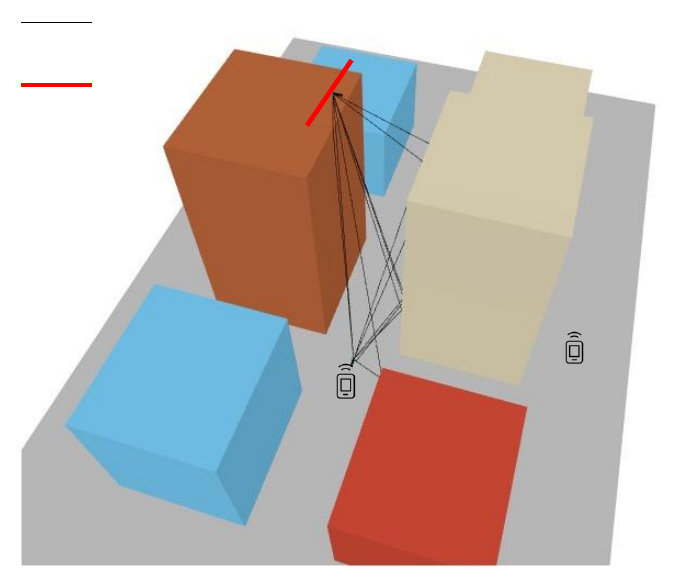}};
    \gettikzxy{(image.north east)}{\ix}{\iy};
    
    \node at (0.3*\ix,0.99*\iy)[rotate=0,anchor=north]{{channel paths}};
    \node at (0.3*\ix,0.88*\iy)[rotate=0,anchor=north]{{antenna array}};
    \node at (0.5*\ix,0.33*\iy)[rotate=0,anchor=north]{{user 1}};
    \node at (0.82*\ix,0.38*\iy)[rotate=0,anchor=north]{{user 2}};
    
    \end{tikzpicture}
    \caption{A street canyon scenario~\cite{sionna}, where channel paths are shown for user 1 for example. Note that user 1 is a \ac{los} user, while user 2 is a \ac{nlos} user as its \ac{los} path is blocked by a building.}
    \label{fig:system_setting}
    \vspace{-3mm}
\end{figure}
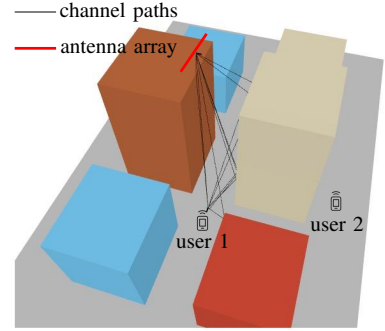

In general, wireless localization can be broadly categorized into two main classes: 1) model-based methods that rely on channel parameter estimation and geometric relationships~\cite{An_Intr_Zeka_2012,Posi_And_Shah_2018,Harn_NLOS_Mend_2019,Robu_Snap_Kalt_2024}; 2) data-driven methods that leverage \ac{ml} to extract user position from \ac{csi}~\cite{Mach_Lear_Sing_2021,Lear_to_Yun_2025,Mach_Lear_Gong_2022}. Specifically, some classical model-based methods perform well in \ac{los} scenarios, e.g., user 1 in Fig.~\ref{fig:system_setting}~\cite{Mult_Emit_Schm_1986,Sign_Reco_Trop_2007,Chan_Esti_Lee_2016,Mode_End_Jose_2025}, but would struggle to keep their superiority in \ac{nlos} scenarios\footnote{In this paper, \ac{nlos} scenarios refer to scenarios where \ac{los} path is blocked. Conversely, \ac{los} scenarios describe cases where the \ac{los} path exists, regardless of whether \ac{nlos} paths are present or not.} due to \ac{los} blockage ( e.g., user 2 in Fig.~\ref{fig:system_setting}). Although more sophisticated methods that utilize \ac{nlos} paths can be adopted, the complexity (especially for channel estimation) is high and their use is usually limited to certain scenarios (e.g., relying on specular reflections)~\cite{Mach_Lear_Sing_2021,Posi_And_Shah_2018,Harn_NLOS_Mend_2019,An_Intr_Zeka_2012,Robu_Snap_Kalt_2024}. On the contrary, data-driven methods (including supervised and unsupervised learning) hold some promise for \ac{nlos} scenarios, but suffer from certain drawbacks. For example, supervised learning, e.g., fingerprinting~\cite{Mach_Lear_Sing_2021,Lear_to_Yun_2025,Mach_Lear_Gong_2022}, requires extensive data measurement and collection for each specific setting, which may not be practical especially for dynamic environments. On the other hand, although unsupervised channel charting can be used for localization, position estimates are highly distorted, and the global geometry cannot be preserved if no additional side-information, e.g., anchor locations, is available~\cite{Chan_Char_Stud_2018, Chan_Char_Sued_2025, Angl_Prof_Step_2024, Trip_Wire_Ferr_2021}. To the best of our knowledge, unsupervised learning for localization remains an open challenge and has not been adequately addressed to date.

In this paper, as model-based and \ac{ml}-based methods have their respective advantages, we aim to combine them to mutually enhance overall positioning performance. In summary, the main contributions of this work are given as follows: 1) A unified localization method is proposed, where \ac{los}/\ac{nlos} identification is conducted, based on which the model-based method or a \ac{nn} model (obtained by unsupervised learning) is adopted for localization; 2) In order to reduce the reliance on exhaustive field measurement campaigns and ground-truth label collection, a \ac{ml}-based method is presented in a manner of unsupervised learning, where self-generated labels are used for \ac{nn} training. By resorting to \ac{ot}~\cite{Opti_Tran_Cour_2017,Comp_Opti_Peyr_2019}, these self-generated training labels are derived jointly from the position estimates of model-based methods and map information; 3) Ray-tracing simulations are carried out to evaluate the unified method, which was shown to outperform model-based methods and achieve a competitive performance with fingerprinting. We are basically bridging the performance gap (e.g., positioning accuracy) between unsupervised channel charting and fully-supervised fingerprinting, by leveraging minimal side-information (i.e., building layouts).

\section{System Model}
We consider a single-cell \ac{ofdm} system, where a \ac{bs} equipped with $M$ antennas serves multiple single-antenna users over $N_c$ subcarriers. A dedicated sequence of pilots is transmitted by each user such that its associated \ac{csi} can be estimated by the \ac{bs}. We assume perfect channel estimation at the \ac{bs}, and the \ac{csi} from user $i$ at all subcarriers is denoted by $\mathbf{H}_i \in \mathbb{C}^{M \times N_c}$. Moreover, tight synchronization between users and the \ac{bs} is achievable through protocols like synchronization signal transmission, \ac{rtt} measurements, and timing advance mechanism~\cite{Phys_Laye_3GPP_2025}. Our goal is to estimate the position of each user, denoted by $\mathbf{p}_i = [ x_i, y_i, z_i]^\mathsf{T} $, based solely on its associated \ac{csi}, i.e., $\hat{\mathbf{p}}_i = f_e(\mathbf{H}_i)$, where $\hat{\mathbf{p}}_i$ is the estimate of $\mathbf{p}_i$ and $f_e(\cdot)$ is an estimation function.

We assume that a 3D map of the environment is accessible, based on which the region for all possible user positions can be extracted and denoted by $\mathcal{R} \subseteq \mathbb{R}^{3 \times 1}$. Therefore, for any user position in the system, we have $\mathbf{p}_i \in \mathcal{R}$, $\forall i$. Moreover, the 3D map enables us to determine whether a given point in $\mathcal{R}$ has a \ac{los} connection to the \ac{bs} or not, based on which the regions for \ac{los} and \ac{nlos} users can be distinguished. We denote the regions for \ac{los} users and \ac{nlos} users as $\mathcal{R}_{\rm LoS} \subseteq \mathbb{R}^{3 \times 1}$ and $\mathcal{R}_{\rm NLoS} \subseteq \mathbb{R}^{3 \times 1}$, respectively, where $\mathcal{R}_{\rm LoS} \cap \mathcal{R}_{\rm NLoS} = \emptyset$ and $\mathcal{R}_{\rm LoS} \cup \mathcal{R}_{\rm NLoS} = \mathcal{R}$.

A \ac{ula} with inter-antenna separation of $d$ is adopted at the \ac{bs}. Considering multi-path effect, the frequency domain channel matrix for user $i$ can be modeled as $\mathbf{H}_i = \sum_{l = 1}^{L_i} \beta_l^i \mathbf{a}_{\mathbf{t}}(\theta_l^i) \mathbf{b}^\mathsf{T}(\tau_l^i)$, $\forall i$~\cite{Mult_ISAC_Visa_2024,Mode_End_Jose_2025}, where there are $L_i$ channel paths for user $i$; $\beta_l^i$, $\theta_l^i$, and $\tau_l^i$ are the complex channel gain, \ac{aoa}, and \ac{toa} of the $l$-th path, respectively; $\mathbf{b}(\tau) \in \mathbb{C}^{N_c \times 1}$ is the frequency-domain steering vector, and $[\mathbf{b}(\tau)]_n = \exp [-\jmath 2 \pi (n-1) \Delta_f \tau]$, $n = 1,2, \cdots,N_c$, with $\Delta_f$ being the subcarrier spacing; $\mathbf{a}_{\mathbf{t}}(\theta) \in \mathbb{C}^{M \times 1}$ is array steering vector at the \ac{bs}, and $[\mathbf{a}_{\mathbf{t}}(\theta)]_m = \exp [\jmath 2 \pi (m-1) \frac{d}{\lambda} \sin (\theta)]$, $m = 1,2, \cdots,M$, where $\lambda = {c}/{f_c}$ with $c$ being the speed of light and $f_c$ being the carrier frequency.

\section{Unified Localization}

\begin{figure}[t]
    \centering
    \begin{tikzpicture}[every node/.style={font=\footnotesize}]
    \node (image) [anchor=south west]{\includegraphics[width=0.86\linewidth]{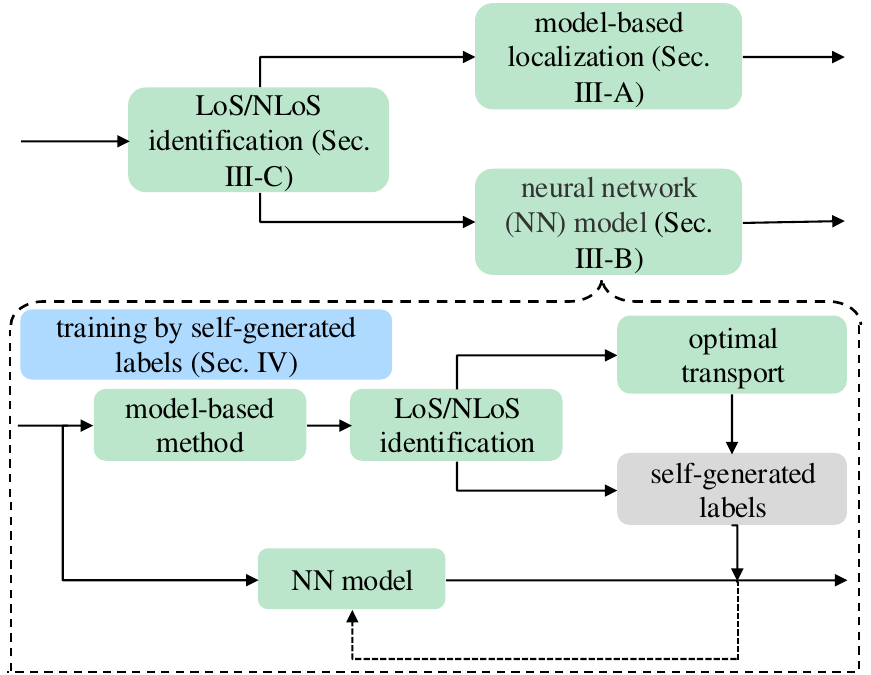}};
    \gettikzxy{(image.north east)}{\ix}{\iy};
    
    \node at (0.09*\ix,0.85*\iy)[rotate=0,anchor=north]{{$\mathbf{H}_i$}};
    \node at (0.4*\ix,0.97*\iy)[rotate=0,anchor=north]{{LoS}};
    \node at (0.4*\ix,0.67*\iy)[rotate=0,anchor=north]{{NLoS}};
    \node at (0.9*\ix,0.98*\iy)[rotate=0,anchor=north]{{$\hat{\mathbf{p}}_i$}};
    \node at (0.9*\ix,0.74*\iy)[rotate=0,anchor=north]{{$\hat{\mathbf{p}}_i$}};

    \node at (0.08*\ix,0.45*\iy)[rotate=0,anchor=north]{{$\{\mathbf{H}_i\}$}};
    \node at (0.6*\ix,0.545*\iy)[rotate=0,anchor=north]{{$I'(\mathbf{H}_i) = 0$}};
    \node at (0.6*\ix,0.29*\iy)[rotate=0,anchor=north]{{$I'(\mathbf{H}_i) = 1$}};
    \node at (0.79*\ix,0.13*\iy)[rotate=0,anchor=north]{{$\{\tilde{\mathbf{p}}_i\}$}};
    \node at (0.92*\ix,0.16*\iy)[rotate=0,anchor=north]{{$\{\hat{\mathbf{p}}_i^{\rm nb}\}$}};
    \node at (0.62*\ix,0.11*\iy)[rotate=0,anchor=north]{{training}};
    
    \end{tikzpicture}
    \caption{The unified localization method, where the dashed border represents the training process, including training label self-generation, of the \ac{nn} model.}
    \label{fig:framework}
\end{figure}
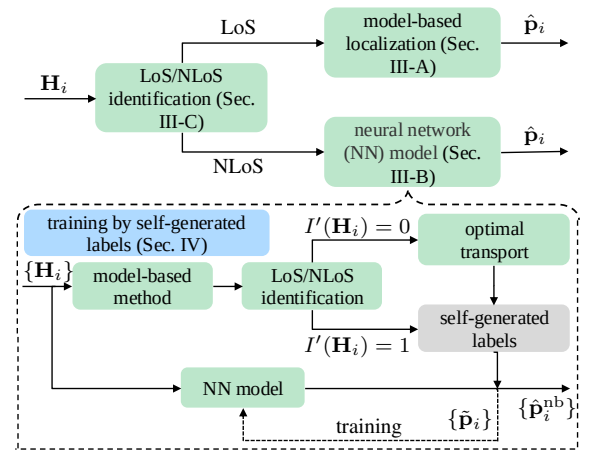

The unified localization method is illustrated in Fig.~\ref{fig:framework}. Firstly, \ac{los}/\ac{nlos} identification is conducted based on \ac{csi} $\mathbf{H}_i$; then, if user $i$ is identified as a \ac{los} user, the model-based method would be used to estimate its position; otherwise, a \ac{nn} model is adopted to generate $\hat{\mathbf{p}}_i$ based on $\mathbf{H}_i$. The model-based and \ac{nn}-based methods are integrated as
\begin{equation}\label{eq:unified_localization}
    \hat{\mathbf{p}}_i = \left\{ \begin{array}{ll}
        f_{\rm pe}\left[ f_{\rm ce}(\mathbf{H}_i) \right], & {\rm if} \, I'(\mathbf{H}_i) = 1;  \\
        f_{\Theta}\left[f_{\rm extr}(\mathbf{H}_i)\right], & {\rm if} \, I'(\mathbf{H}_i) = 0,
    \end{array} \right.
\end{equation}
where $f_{\rm ce}(\cdot)$ is a channel parameter estimation function; $f_{\rm pe}(\cdot)$ is a function that maps channel parameters to position based on geometric relationships; $f_{\rm extr}(\cdot)$ is a channel feature extraction function to transform \ac{csi} $\mathbf{H}_i$ to the input of the \ac{nn} model; $f_{\Theta}(\cdot)$ is the forward function of the \ac{nn} model with $\Theta$ being parameters; and $I'(\cdot)$ is a \ac{csi}-based \ac{los}/\ac{nlos} identification function. Note that the \ac{nn} model is obtained by unsupervised learning, where self-generated training labels $\{\tilde{\mathbf{p}}_i\}$ are used to learn the relationship between \ac{csi} and user position. More details about each block are given as follows.

\subsection{Model-based Method}
With \ac{ofdm} transmission and multi-antenna \ac{bs}, channel parameters, including channel gain, \ac{aoa}, and \ac{toa} for different channel paths, can be estimated by channel estimation algorithms, e.g., \ac{music} algorithm~\cite{Mult_Emit_Schm_1986}, based on which the position of a user can be derived by utilizing geometric relationships\cite{An_Intr_Zeka_2012,Mult_ISAC_Visa_2024,Mode_End_Jose_2025}. In this paper, the \ac{omp} algorithm~\cite{Sign_Reco_Trop_2007,Chan_Esti_Lee_2016,Mode_End_Jose_2025} is used, where $\hat{L}_i$ channel paths can be recovered for user $i$ with channel parameters $\{ \hat{\beta}_l^i, \hat{\theta}_l^i, \hat{\tau}_l^i \}$, $l = 1,2,\ldots,\hat{L}_i$ (for more details, the reader is referred to Algorithm 1 of~\cite{Mode_End_Jose_2025}). For a low complexity implementation, the shortest channel path among the recovered $\hat{L}_i$ paths is identified as the \ac{los} path and used to estimate the user position. Thus, the channel parameter estimation function $f_{\rm ce}(\cdot)$ that maps the \ac{csi} to \ac{aoa} and \ac{toa} can be written as $\{ \hat{\theta}_s^i, \hat{\tau}_s^i \} = f_{\rm ce}(\mathbf{H}_i)$, $\forall i$, where $s = \arg \min_l \hat{\tau}_l^i$, $l = 1,2,\ldots,\hat{L}_i$.

With $\{ \hat{\theta}_s^i, \hat{\tau}_s^i \}$ (yielding sufficient information to estimate x-coordinate and y-coordinate for a given z-coordinate), user position estimate can be derived by regarding this channel path as \ac{los}, based on the following geometric relationships:
\begin{equation}\label{eq:geometric_relationships}
    \hat{\tau}_s^i = \frac{\| \mathbf{p}_{\rm BS} - \hat{\mathbf{p}}_i \|_2}{c}, \quad \hat{\theta}_s^i = \arcsin \left( \frac{(\hat{\mathbf{p}}_i - \mathbf{p}_{\rm BS}) \cdot \mathbf{n}_{\rm ULA}}{\|\hat{\mathbf{p}}_i - \mathbf{p}_{\rm BS} \|_2} \right),
\end{equation}
where $\mathbf{p}_{\rm BS}$ is the position of the \ac{bs} and $\mathbf{n}_{\rm ULA}$ is the normalized vector representing the direction of the \ac{ula}. Therefore, the model-based estimate $\hat{\mathbf{p}}_i^{\rm mb}$ can be written as $\hat{\mathbf{p}}_i^{\rm mb} = f_{\rm pe}\left[ f_{\rm ce}(\mathbf{H}_i) \right]$, $\forall i$, where $f_{\rm pe}(\cdot)$ can be obtained, for example, by solving~\eqref{eq:geometric_relationships} (see, e.g.,~\cite{A_Tuto_Chen_2022,An_Intr_Zeka_2012,Posi_And_Shah_2018,Harn_NLOS_Mend_2019,Robu_Snap_Kalt_2024} for more details on geometric relationships).

\subsection{Unsupervised Learning Method}
In order to obtain a parametric model, \ac{nn} is adopted to map \ac{csi} to position. Instead of directly feeding $\mathbf{H}_i$ to \ac{nn} model, \ac{csi} preprocessing is used to extract features, which not only reduces the dimension of input data but also makes \ac{nn} easier to learn hidden patterns~\cite{Lear_to_Yun_2025,Mach_Lear_Gong_2022}. To do so, we first convert the antenna-frequency domain \ac{csi} $\mathbf{H}_i$ to the angle-delay domain, as given by $\bar{\mathbf{H}}_i = \mathcal{F}_a\left[ \mathcal{F}^{-1}_b (\mathbf{H}_i)\right]$, $\forall i$, where $\mathcal{F}_a(\cdot)$ is the discrete Fourier transform along the antenna axis; $\mathcal{F}^{-1}_b(\cdot)$ is the inverse discrete Fourier transform along the frequency axis.

Since all channel paths in $\mathbf{H}_i$ have $\tau_l^i \in [\tau_{\min}, \tau_{\max}]$, where $\tau_{\min}$ and $\tau_{\max}$ are the minimum \ac{toa} and maximum \ac{toa}, the angle-delay domain \ac{csi} $\bar{\mathbf{H}}_i$ can be truncated by taking the columns that have delays within $\tau_{\min}$ and $\tau_{\max}$. Specifically, we truncate $\bar{\mathbf{H}}_i$ as $\bar{\mathbf{H}}_i^{\rm t} = [\bar{\mathbf{H}}_i]_{:,\left \lfloor{\tau_{\min} \Delta_f N_c}\right \rfloor :\left \lceil \tau_{\max} \Delta_f N_c\right \rceil }$, $\forall i$, whose element-wise amplitude and phase can be used to generate the input to the \ac{nn}\footnote{In this paper, since ideal \ac{csi} is assumed, channel amplitude and phase are both utilized as the input to the \ac{nn} model. However, with real datasets, e.g., including \ac{csi} from multiple \ac{bs}s without synchronization and/or from users with different transmit powers, channel feature extraction could be adapted accordingly, for example, by averaging over a fraction of time, normalizing the channel and/or removing the phase information~\cite{Chan_Char_Sued_2025}.}, as denoted by $\mathbf{d}_i = {\rm vec}\left( \left[\log (|\bar{\mathbf{H}}_i^{\rm t}|), \angle \bar{\mathbf{H}}_i^{\rm t} \right] \right)$, $\forall i$, where ${\rm vec} (\cdot)$ would vectorize a matrix. Thus, implementing \ac{csi} domain conversion, truncation,  and element-wise extraction, $f_{\rm extr}(\cdot)$ transforms $\mathbf{H}_i$ to the input of the \ac{nn} model, i.e., $\mathbf{d}_i = f_{\rm extr}(\mathbf{H}_i)$, based on which the output of the \ac{nn} model, i.e., the user position estimate $\hat{\mathbf{p}}_i^{\rm nb}$, can be expressed by $\hat{\mathbf{p}}_i^{\rm nb} = f_{\Theta}[f_{\rm extr}(\mathbf{H}_i)]$, $\forall i$.

Given a training dataset $\mathcal{S}: \{\mathbf{H}_i\}_{i = 1,2,\ldots,N_u}$, where $N_u$ is the number of user positions, an extended dataset $\mathcal{S}': \{\mathbf{d}_i,\tilde{\mathbf{p}}_i\}_{i = 1,2,\ldots,N_u}$ can be obtained by applying $f_{\rm extr}(\cdot)$ and training label self-generation. The self-generated labels $\{\tilde{\mathbf{p}}_i\}$ can be obtained based on \ac{csi} $\{\mathbf{H}_i\}$ and \ac{ot} as shown in Fig.~\ref{fig:framework}: for identified \ac{los} users, $\{\tilde{\mathbf{p}}_i|I'(\mathbf{H}_i) = 1\}$ are the model-based estimates $\{\hat{\mathbf{p}}_i^{\rm mb}|I'(\mathbf{H}_i) = 1\}$; while for identified \ac{nlos} users, $\{\tilde{\mathbf{p}}_i|I'(\mathbf{H}_i) = 0\}$ are obtained by transforming $\{\hat{\mathbf{p}}_i^{\rm mb}|I'(\mathbf{H}_i) = 0\}$ via \ac{ot}. More details are deferred to Sec.~\ref{sec:OT}. Then, the loss function of the \ac{nn} model can be designed as
\begin{equation}\label{eq:loss_func}
    \mathcal{L} = \frac{1}{N_u} \sum_{\mathbf{d}_i, \tilde{\mathbf{p}}_i \in \mathcal{S}'} \| f_{\Theta}(\mathbf{d}_i) - \tilde{\mathbf{p}}_i \|^2_2,
\end{equation}
based on which the parameters $\Theta$ can be updated by backpropagation during training with the labels $\{\tilde{\mathbf{p}}_i\}$ that are self-generated before. As in Sec.~\ref{sec:results}, when an $L_{\rm MLP}$-layer \ac{mlp} with layer widths $\{n_l\}_{l=1, \ldots,L_{\rm MLP}}$ is adopted for the \ac{nn} model, the complexity of training is $\mathcal{O}\left(N_u\sum_{l=1}^{L_{\rm MLP}} n_{l-1} n_l \right)$ per epoch, where $n_0$ is the dimensionality of $\mathbf{d}_i$.

\subsection{LoS/NLoS Identification}
\ac{csi}-based \ac{los}/\ac{nlos} identification has been widely studied and numerous algorithms are proposed, where some of them achieve a high identification accuracy, e.g., $93\% \text{-} 100\%$ for different system settings~\cite{An_Intr_Zeka_2012,NLOS_Dete_Schr_2007}. It is assumed that we have a \ac{los}/\ac{nlos} identification mechanism, denoted by $I(\cdot)$, where user $i$ is identified as a \ac{los} user if $I(\mathbf{H}_i)=1$ and a \ac{nlos} user if $I(\mathbf{H}_i)=0$. For the given $I(\cdot)$, the identification accuracy is denoted by $p_I \in [0.5,1]$. Furthermore, with a \ac{los} path, the mapping function $f_{\rm pe}(\cdot)$, e.g., solving~\eqref{eq:geometric_relationships} based on \ac{aoa} and \ac{toa}, would result in a position in the \ac{los} region. This allows us to conclude that all \ac{los} users would be localized in the \ac{los} region with $f_{\rm pe}\left[ f_{\rm ce}(\cdot) \right]$, i.e., if a user is localized in the \ac{nlos} region, it must be a \ac{nlos} user\footnote{However, \ac{nlos} users could be localized in \ac{los} region by $f_{\rm pe}\left[ f_{\rm ce}(\cdot) \right]$ due to reflection, diffraction, or scattering of the physical environment.}. Therefore, given the results from $I(\mathbf{H}_i)$, we can further improve the \ac{los}/\ac{nlos} identification by
\begin{equation}\label{eq:LoS_identification}
    I'(\mathbf{H}_i) = \left\{ \begin{array}{ll}
        0, & {\rm if} \, f_{\rm pe}\left[ f_{\rm ce}(\mathbf{H}_i) \right] \in \mathcal{R}_{\rm NLoS};  \\
        I(\mathbf{H}_i), & {\rm if} \, f_{\rm pe}\left[ f_{\rm ce}(\mathbf{H}_i) \right] \in \mathcal{R}_{\rm LoS}.
    \end{array} \right.
\end{equation}

\begin{remark}
    When the mechanism $I(\cdot)$ is not accessible or its identification accuracy $p_I$ drops to a extremely low value, a conservative method that does not rely on $I(\cdot)$ can be presented. In particular, without $I(\cdot)$, the model-based results could be used for \ac{los}/\ac{nlos} identification, i.e., if a user is localized in the \ac{los} (resp. \ac{nlos}) region based on $f_{\rm pe}\left[ f_{\rm ce}(\cdot) \right]$, it would be identified as a \ac{los} (resp. \ac{nlos}) user. Therefore, the \ac{los}/\ac{nlos} identification~\eqref{eq:LoS_identification} can be reformulated as
\begin{equation}\label{eq:LoS_identification_conservative}
    I'(\mathbf{H}_i) = \left\{ \begin{array}{ll}
        0, & {\rm if} \, f_{\rm pe}\left[ f_{\rm ce}(\mathbf{H}_i) \right] \in \mathcal{R}_{\rm NLoS};  \\
        1, & {\rm if} \, f_{\rm pe}\left[ f_{\rm ce}(\mathbf{H}_i) \right] \in \mathcal{R}_{\rm LoS},
    \end{array} \right.
\end{equation}
based on which \ac{los} and \ac{nlos} users can be identified completely by $f_{\rm pe}\left[ f_{\rm ce}(\mathbf{H}_i) \right]$. Then, referring to~\eqref{eq:unified_localization}, the position estimates for identified \ac{los} (resp. \ac{nlos}) users are obtained by $\hat{\mathbf{p}}_i = f_{\rm pe}\left[ f_{\rm ce}(\mathbf{H}_i) \right]$ (resp. $\hat{\mathbf{p}}_i = f_{\Theta}\left[f_{\rm extr}(\mathbf{H}_i)\right]$).
\end{remark}

\section{\ac{ot}-based Label Self-Generation}\label{sec:OT}

\subsection{Discrete Optimal Transport}
Given a source domain $\mathcal{X}_s$ with a defined probability measure $u_s$ and a target domain $\mathcal{X}_t$ with a defined probability measure $u_t$, \ac{ot} aims to find a transformation from $\mathcal{X}_s$ to $\mathcal{X}_t$, as denoted by $\mathbf{T}^*(\cdot)$, that minimizes the transportation cost on the condition that the image measure of $u_s$ in $\mathcal{X}_t$, as denoted by $u_s'$ (since $\mathbf{T}^*(\cdot)$ transforms $u_s$ to $u_s'$, we have $u_s(\mathbf{x}_s) =  u_s'[\mathbf{T}^*(\mathbf{x}_s)]$, $\forall \mathbf{x}_s \in \mathcal{X}_s$), is $u_s'(\mathbf{x}_t) =  u_t(\mathbf{x}_t)$, $\forall \mathbf{x}_t \in \mathcal{X}_t$.
Therefore, the \ac{ot} problem can be written as~\cite{Opti_Tran_Cour_2017}
\begin{equation}\label{eq:OT_problem}
\begin{aligned}
    \mathbf{T}^* =  \arg \min_{\mathbf{T}} & \quad \int_{\mathcal{X}_s} c[\mathbf{x}_s,\mathbf{T}(\mathbf{x}_s)]\text{d}u_s(\mathbf{x}_s), \\
    {\rm s.t.} & \quad u_s(\mathbf{x}_s) =  u_t[\mathbf{T}(\mathbf{x}_s)],\, \forall \mathbf{x}_s \in \mathcal{X}_s,
\end{aligned}
\end{equation}
where $c[\mathbf{x}_s,\mathbf{T}(\mathbf{x}_s)]$ is the transportation cost function to transform $\mathbf{x}_s$ to $\mathbf{T}(\mathbf{x}_s)$.

The convex relaxation of the above problem~\eqref{eq:OT_problem} can be formulated as finding a joint probability function $\gamma$ with marginals $u_s$ and $u_t$ to minimize $\int_{\mathcal{X}_s \times \mathcal{X}_t} c(\mathbf{x}_s,\mathbf{x}_t)d\gamma(\mathbf{x}_s,\mathbf{x}_t)$~\cite{Comp_Opti_Peyr_2019}. Moreover, when $\mathcal{X}_s = \{\mathbf{x}_s^i\}_{i = 1,2,\ldots,N_s}$ and $\mathcal{X}_t = \{\mathbf{x}_t^i\}_{i = 1,2,\ldots,N_t}$, where $N_s$ and $N_t$ are the number of data samples in $\mathcal{X}_s$ and $\mathcal{X}_t$, the probability measures $u_s$ and $u_t$ can be given in the form of $\mathbf{u}_s$ and $\mathbf{u}_t$, which are the probability vectors of the data samples in $\mathcal{X}_s$ and $\mathcal{X}_t$, respectively. Therefore, optimizing the joint probability matrix $\mathbf{\Gamma} \in \mathbb{R}_+^{N_s \times N_t}$ across $\mathcal{X}_s$ and $\mathcal{X}_t$, the problem~\eqref{eq:OT_problem} in discrete case can be approximated as~\cite{Comp_Opti_Peyr_2019}
\begin{equation}\label{eq:OT_opti}
    \mathbf{\Gamma}^* = \arg \min_\mathbf{\Gamma \in \mathcal{B}} \quad \langle\mathbf{\Gamma}, \mathbf{C}\rangle_F,
\end{equation}
where $\langle \cdot, \cdot\rangle_F$ is the Frobenius dot product, $\mathbf{C} \in \mathbb{R}_+^{N_s \times N_t}$ is the transportation cost matrix with $[\mathbf{C}]_{i,j} = c(\mathbf{x}_s^i,\mathbf{x}_t^j)$, and
\begin{equation}
    \mathcal{B} = \{ \mathbf{\Gamma} \in \mathbb{R}_+^{N_s \times N_t} | \mathbf{\Gamma} \cdot \mathbf{1}_{N_s} = \mathbf{u}_s, \,\, \mathbf{\Gamma}^{\mathsf{T}} \cdot \mathbf{1}_{N_t} = \mathbf{u}_t\},
\end{equation}
with $\mathbf{1}_N$ being a all-one vector with the size of $N$. Note that $\mathcal{B}$ represents all possible joint probability matrices $\{\mathbf{\Gamma}\}$ that are with marginals $\mathbf{u}_s$ and $\mathbf{u}_t$. It is observed that the discrete \ac{ot} problem~\eqref{eq:OT_opti} is a linear programming and can be solved efficiently by, e.g., simplex methods~\cite{Opti_Tran_Cour_2017}, which usually infer a polynomial time complexity. Moreover, low-complexity algorithms can be adopted to approximately solve~\eqref{eq:OT_opti}, e.g., the Sinkhorn algorithm~\cite{Opti_Tran_Cour_2017} with a complexity of $\mathcal{O}\left( I_{\rm Iter} N_s N_t \right)$, where $I_{\rm Iter}$ is the number of Sinkhorn iterations.

\subsection{Self-generated Training Labels}
Given the training dataset $\mathcal{S}: \{\mathbf{H}_i\}_{i = 1,2,\ldots,N_u}$, position estimates $\{\hat{\mathbf{p}}_i^{\rm mb}\}_{i = 1,2,\ldots,N_u}$ can be attained based on the model-based method $\hat{\mathbf{p}}_i^{\rm mb} = f_{\rm pe}\left[ f_{\rm ce}(\mathbf{H}_i) \right]$. For identified \ac{los} users, $\{\hat{\mathbf{p}}_i^{\rm mb}\}$ can be used as training labels, i.e., $\tilde{\mathbf{p}}_i = \hat{\mathbf{p}}_i^{\rm mb}$ if $I'(\mathbf{H}_i) = 1$; however, these estimates $\hat{\mathbf{p}}_i^{\rm mb}$ may not be accurate for identified \ac{nlos} users and do not match the map. Therefore, the next task is to find a transformation $\mathbf{T}^*(\cdot)$ that maps $\{\hat{\mathbf{p}}_i^{\rm mb}|I'(\mathbf{H}_i) = 0\}_{i = 1,2,\ldots,N_u}$ to some other positions that can match the \ac{nlos} region on the map.

To achieve this goal, we resort to the discrete \ac{ot}~\eqref{eq:OT_opti}, where the source domain $\mathcal{X}_s = \{\hat{\mathbf{p}}_i^{\rm mb}|I'(\mathbf{H}_i) = 0\}_{i = 1,2,\ldots,N_u}$ and the target domain $\mathcal{X}_t = \{\bar{\mathbf{p}}_i\}_{i = 1,2,\ldots,N_t}$, based on which the joint probability matrix $\mathbf{\Gamma} \in \mathbb{R}_+^{N_s \times N_t}$ is optimized ($N_s = N_u-\sum_i I'(\mathbf{H}_i)$). In this paper, the cost associated with two positions is set as the squared Euclidean distance\footnote{\label{footnote:index_ordering}Note that the ${i'}$-th data sample in $\{\hat{\mathbf{p}}_i^{\rm mb}|I'(\mathbf{H}_i) = 0\}_{i = 1,2,\ldots,N_u}$ is the $i$-th data sample in $\{\hat{\mathbf{p}}_i^{\rm mb}\}_{i = 1,2,\ldots,N_u}$.}, i.e., $[\mathbf{C}]_{i',j} = \|\hat{\mathbf{p}}_{i}^{\rm mb} - \bar{\mathbf{p}}_j \|_2^2$. Based on the map, the data samples in the target domain $\{\bar{\mathbf{p}}_i\}_{i = 1,2,\ldots,N_t}$ can be generated by selecting $N_t$ grid points in the \ac{nlos} region, i.e., $\bar{\mathbf{p}}_i \in \mathcal{G}_{\rm NLoS} \subseteq \mathcal{R}_{\rm NLoS}$, where $\mathcal{G}_{\rm NLoS}$ contains all grid points with a spacing distance of $\Delta_d$ in $\mathcal{R}_{\rm NLoS}$. If all users are located uniformly on the map\footnote{Note that the assumption of uniformly distributed users on the map is not mandatory; on the contrary, any distribution is allowed in general for \ac{ot}.}, the probability vectors $\mathbf{u}_s$ and $\mathbf{u}_t$ can be written as
\begin{equation}
    \mathbf{u}_s = \frac{1}{N_s} \cdot \mathbf{1}_{N_s}, \, \, \mathbf{u}_t = \frac{1}{N_t} \cdot \mathbf{1}_{N_t}.
\end{equation}

Once the optimal joint probability matrix $\mathbf{\Gamma}^*$ is obtained, the transformed position for each identified \ac{nlos} user can be expressed by~\cite{Opti_Tran_Cour_2017}
\begin{equation}
    \tilde{\mathbf{{P}}}_{\rm NLoS} = {\rm diag}(\mathbf{\Gamma}^* \cdot \mathbf{1}_{N_t})^{-1} \cdot \mathbf{\Gamma}^* \cdot \bar{\mathbf{P}},
\end{equation}
where $\bar{\mathbf{P}} \in \mathbb{R}^{N_t \times 3}$ is the collection of all $\{\bar{\mathbf{p}}_i\}_{i = 1,2,\ldots,N_t}$; $\tilde{\mathbf{{P}}}_{\rm NLoS} \in \mathbb{R}^{N_s \times 3}$ with $[\tilde{\mathbf{{P}}}_{\rm NLoS}]_{i',:} = \mathbf{T}^*(\hat{\mathbf{p}}_{i}^{\rm mb})$ being the mapped position\footref{footnote:index_ordering}. Then, combining the identified \ac{los} and \ac{nlos} users, the self-generated labels $\{\tilde{\mathbf{p}}_i\}_{i = 1,2,\ldots,N_u}$ can expressed as
\begin{equation}
    \tilde{\mathbf{p}}_i = \left\{ \begin{array}{ll}
        \hat{\mathbf{p}}_i^{\rm mb}, & {\rm if} \, I'(\mathbf{H}_i) = 1;  \\
        \mathbf{T}^*(\hat{\mathbf{p}}_i^{\rm mb}), & {\rm if} \, I'(\mathbf{H}_i) = 0.
    \end{array} \right.
\end{equation}

\section{Numerical Experiments}\label{sec:results}

\begin{figure*}[t]
    \centering
    \begin{subfigure}{0.26\textwidth}
        \includegraphics[width=\linewidth]{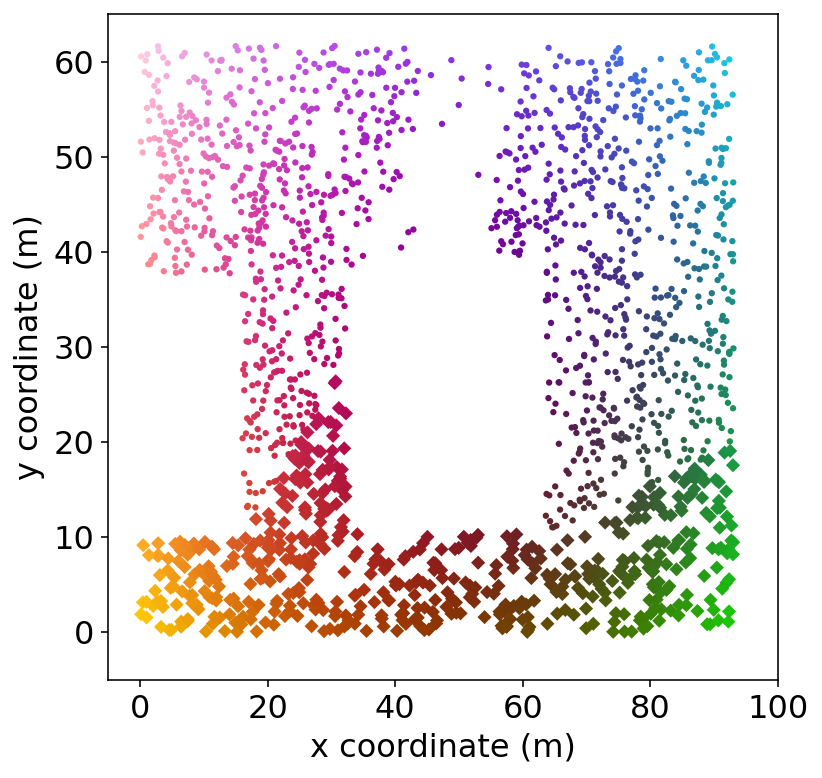}
        \caption{ground-truth positions}
    \end{subfigure}
    \begin{subfigure}{0.26\textwidth}
        \includegraphics[width=\linewidth]{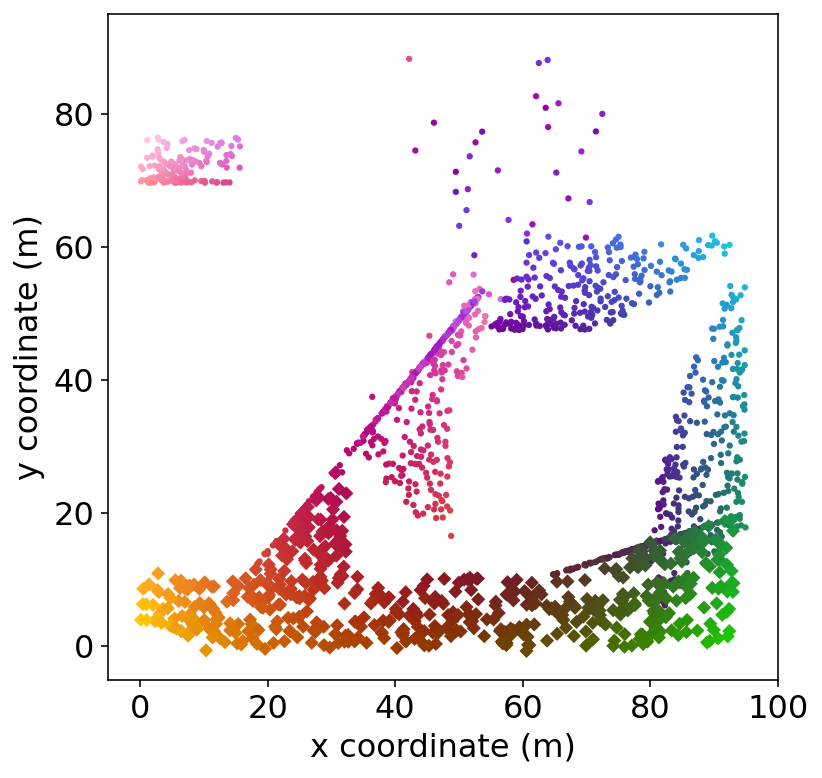}
        \caption{model-based}
    \end{subfigure}
    \begin{subfigure}{0.26\textwidth}
        \includegraphics[width=\linewidth]{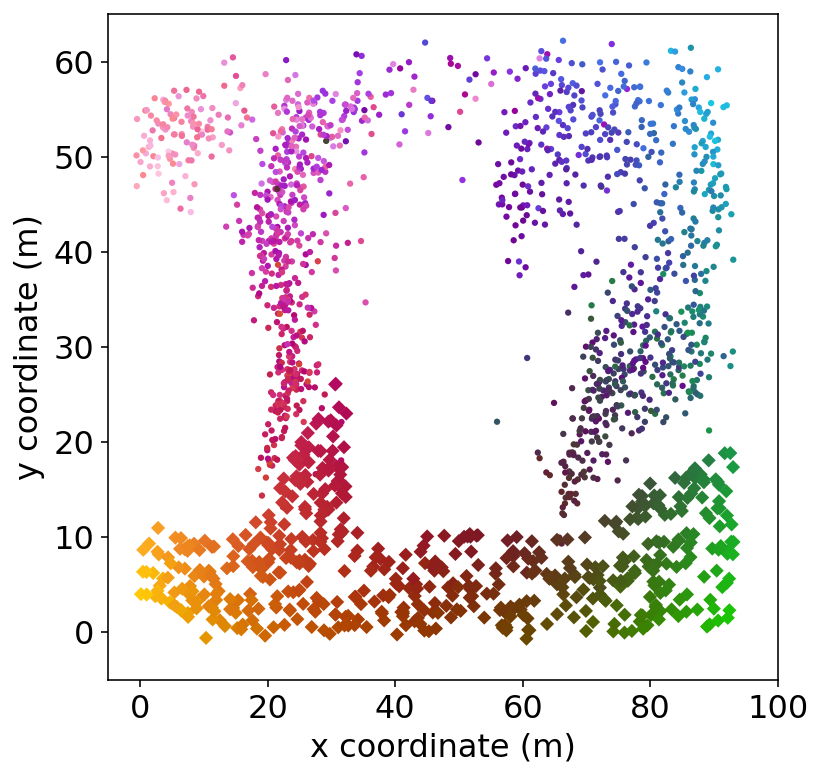}
        \caption{unified method ($p_I = 1$)}
    \end{subfigure}
    \begin{subfigure}{0.26\textwidth}
        \includegraphics[width=\linewidth]{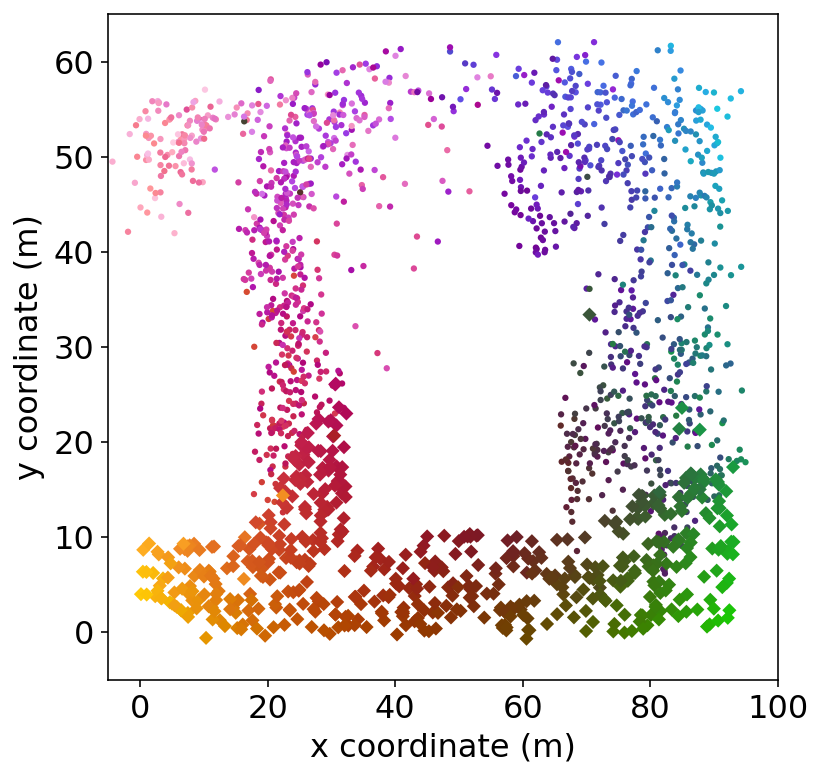}
        \caption{conservative method}
    \end{subfigure}
    \begin{subfigure}{0.26\textwidth}
        \includegraphics[width=\linewidth]{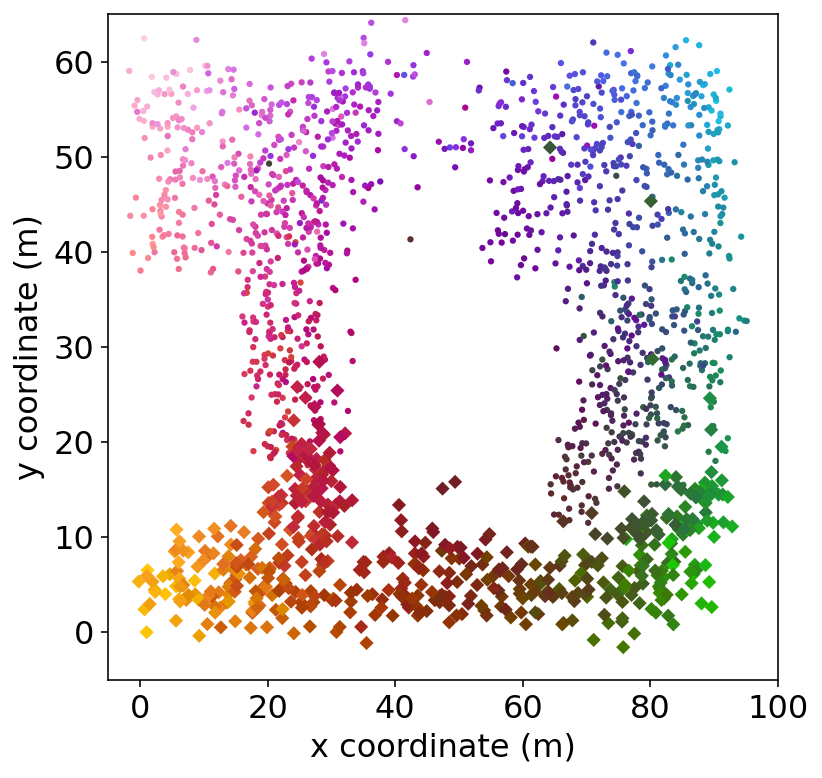}
        \caption{fingerprinting ($\Delta_s = 0.5\, {\rm m}$)}
    \end{subfigure}
    \begin{subfigure}{0.26\textwidth}
        \includegraphics[width=\linewidth]{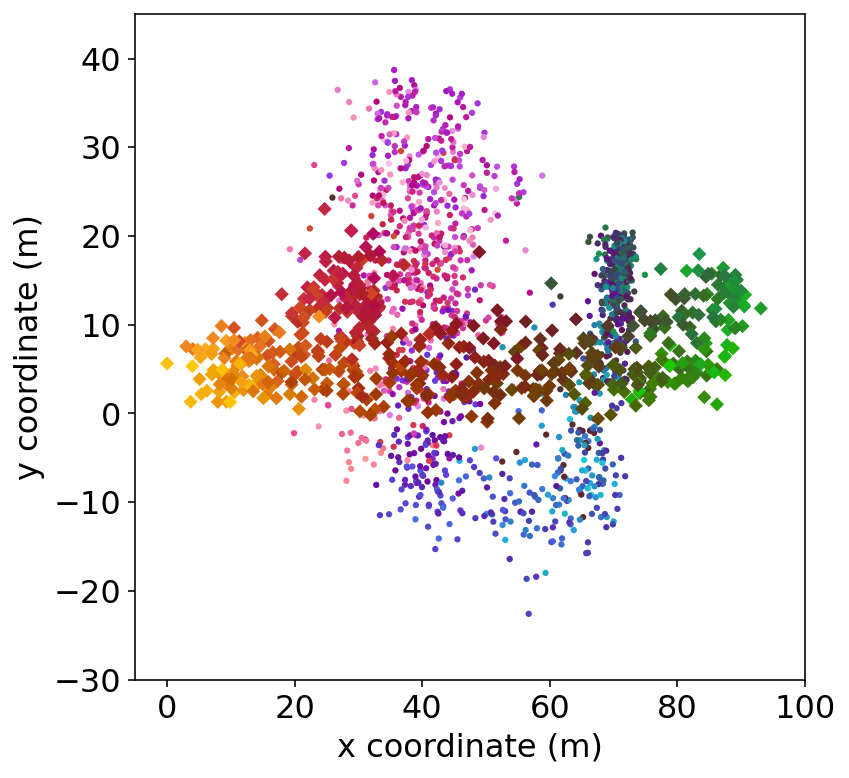}
        \caption{channel charting ($p_I = 1$)}
    \end{subfigure}
    \caption{Position estimates for different localization methods (testing dataset; every user is associated with a unique color; dot marks represent \ac{nlos} users and square marks represent \ac{los} users).}
    \label{fig:estimation_results}
    \vspace{-3mm}
\end{figure*}

\subsection{System Setting}

In this section, numerical experiments in a street canyon scenario, as shown in Fig.~\ref{fig:system_setting}, are carried out to evaluate the unified localization method. The \ac{bs} is placed at $\mathbf{p}_{\rm BS} = [0,-9,57]^{\mathsf{T}}$ with $M=256$ antennas (the inter-antenna distance is set as $d = \lambda/2$), and the height of each user is set as $1.5\, {\rm m}$, i.e.,  $\mathbf{p}_i = [ x_i, y_i, 1.5]^\mathsf{T} $. The carrier frequency of the system is $10\, {\rm GHz}$, and $50\, {\rm MHz}$ bandwidth is used with subcarrier spacing of $120\, {\rm kHz}$. For a better illustration, users are placed randomly in the first quadrant of the map, i.e., $\mathbf{p}_i \in \mathcal{R}$ and $\mathbf{p}_i \in \mathbb{R}_+^{3 \times 1}$, $\forall i$. For each user, the \ac{csi} $\mathbf{H}_i$ is generated realistically by Sionna library~\cite{sionna}, where ray-tracing techniques are used to generate the propagation paths, based on which the channel coefficients are computed. When using \ac{ot}, $N_t$ is set as the total number of all grid points in $\mathcal{G}_{\rm NLoS}$ with $\Delta_d = 0.5\, {\rm m}$.

As for the \ac{ml}-based methods, following the same \ac{nn} architecture and hyperparameters as in~\cite{Trip_Wire_Ferr_2021,Angl_Prof_Step_2024}, a \ac{mlp} is adopted for $f_{\Theta}(\cdot)$, which consists of 5 hidden layers (each with 1024, 512, 256, 128, 64 neurons, respectively, ReLU activation, and batch normalization) and an output layer (with 2 neurons and linear activation). Around 1800 user positions, consisting of 550 \ac{los} and 1250 \ac{nlos}, are generated both for training and testing datasets. Moreover, Adam optimizer with a decayed learning rate is employed for training (1600 epochs).

\subsection{Benchmarks}
In this section, \ac{mae}~\cite{Mach_Lear_Sing_2021,Angl_Prof_Step_2024,Chan_Char_Sued_2025}, i.e., $\frac{1}{N_u}\sum_i \| \hat{\mathbf{p}}_i - \mathbf{p}_i \|_2$, is adopted as a metric for evaluation and analysis. For comparison, in addition to the unified method~\eqref{eq:unified_localization} and its conservative variant, model-based, fingerprinting, and channel charting methods are considered as benchmarks:

\begin{itemize}
    \item \textbf{Model-based method}: position estimates are obtained based on the \ac{omp} algorithm and geometric relationships, i.e., $\hat{\mathbf{p}}_i^{\rm mb} = f_{\rm pe}\left[ f_{\rm ce}(\mathbf{H}_i) \right]$, $\forall i$;
    \item \textbf{Fingerprinting}: position estimates are obtained from \ac{nn}, i.e., $\hat{\mathbf{p}}_i^{\rm nb} = f_{\Theta}\left[f_{\rm extr}(\mathbf{H}_i)\right]$, where $f_{\Theta}(\cdot)$ is trained by assuming ground-truth labels, i.e., $\tilde{\mathbf{p}}_i = \mathbf{p}_i$, $\forall i$. It is a fully-supervised learning approach;
    \item \textbf{Channel charting}\footnote{Channel charting is, in general, a self-supervised learning technique that aims to generate a mapping from high-dimensional \ac{csi} to a low-dimensional
    space, which is called as channel chart, where nearby points correspond to nearby locations in real space. More details are referred to~\cite{Chan_Char_Stud_2018,Angl_Prof_Step_2024,Chan_Char_Sued_2025,Trip_Wire_Ferr_2021}. In this paper, since precise position estimates can be obtained by model-based methods for \ac{los} users, a loss term corresponding to the estimation accuracy of the \ac{los} users is added.}: position estimates are obtained from \ac{nn}, i.e., $\hat{\mathbf{p}}_i^{\rm nb} = f_{\Theta}\left[f_{\rm extr}(\mathbf{H}_i)\right]$, where $f_{\Theta}(\cdot)$ is trained based on the following loss function: $\mathcal{L}_{\rm CC} = \frac{1}{N_u^2}\sum_{\mathbf{H}_i \in \mathcal{S}} \sum_{\mathbf{H}_j \in \mathcal{S}} \frac{\left( \| f_{\Theta}\left(\mathbf{d}_i\right) - f_{\Theta}\left(\mathbf{d}_j\right) \|_2 - \varepsilon d_{i,j} \right) ^2}{d_{i,j}} + \frac{1}{\sum_k I'(\mathbf{H}_k)} \sum_{\{\mathbf{H}_k | I'(\mathbf{H}_k)=1\}} \| f_{\Theta}\left(\mathbf{d}_k\right) - \hat{\mathbf{p}}_k^{\rm mb} \|^2_2$, where $\varepsilon$ is a scaling parameter (learned by training), and $d_{i,j}$ is the feature ``distance" between two \ac{csi}s that would be preserved. In this paper, cosine dissimilarity is used for $\{d_{i,j}\}$~\cite{Angl_Prof_Step_2024}. It is a fully unsupervised learning approach.
\end{itemize}

For a fair comparison, as fingerprinting requires extensive \ac{csi} measurement and collection, the dataset often contains limited data samples. For fingerprinting, grid positions on the map with corresponding \ac{csi} are used for training. Specifically, around 1800 user positions are selected from $\mathcal{G}$, which contains all grid points with a spacing distance of $\Delta_s$ on the map, i.e., $\mathbf{p}_i \in \mathcal{G} \subseteq \mathcal{R}$, $\forall i$. Note that if $\mathcal{G}$ has less than 1800 positions, all grid positions in $\mathcal{G}$ are selected. While for testing, user positions are placed randomly, which is the same as other considered methods.

\subsection{Numerical Results}

\subsubsection{Comparison with benchmarks}

The qualitative results (position estimates) obtained from different localization methods for the testing dataset are shown in Fig.~\ref{fig:estimation_results}. It is observed that the model-based method achieves a high positioning accuracy for \ac{los} users (with a \ac{mae} of 0.34 m), while the estimates of \ac{nlos} users are distorted significantly. The fingerprinting method keeps the global shape of \ac{nlos} users, while the estimates of \ac{los} users are not as precise as the model-based method. Being an integrated strategy, the unified method (also the conservative variant) can not only attain a precise estimation for \ac{los} users, but also the global position of the \ac{nlos} users can be preserved, which shows the effectiveness of the proposed \ac{ot}-based approach in Sec.~\ref{sec:OT} in mapping the \ac{nlos} users to feasible locations that respect the map constraints. Moreover, although the local geometry, i.e., the neighboring relationship between users, is preserved, channel charting cannot capture the global geometry of \ac{nlos} users due to the lack of global coordinate information, resulting in a high estimation error for \ac{nlos} users. This indicates that the model-based methods could provide important information for unsupervised learning methods to improve the estimation accuracy, a strategy followed by the proposed unified method.

\begin{figure}[t]
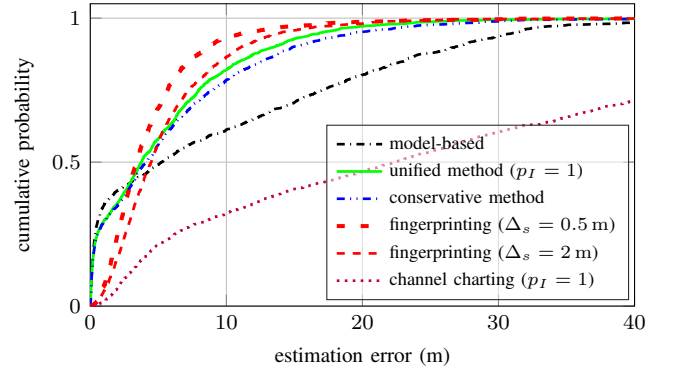

\centering
\begin{minipage}[t]{0.97\linewidth}
\centering
    \include{fig_cdf_com.tex}
    \vspace{-1 cm}
\end{minipage}
\caption{The \ac{cdf} w.r.t.~the positioning error for different methods (testing dataset).}
\label{fig:cdf}
\end{figure}

\begin{table}[t]
\caption{\ac{mae} for different methods (testing dataset).}
\label{tab:MAE}
\centering
\renewcommand{\arraystretch}{1.2} 
\begin{tabular}{|l|c|c|c|}
\hline
\textbf{\ac{mae} (m)} & \textbf{\ac{los}} & \textbf{\ac{nlos}} & \textbf{all users} \\
\hline
\multicolumn{4}{|l|}{\textbf{Unsupervised Methods}} \\
\hline
model-based & \textbf{0.34} & 13.79 & 9.81 \\
\hline
unified method ($p_I = 1$) & \textbf{0.34} & \textbf{7.67} & \textbf{5.50} \\
\hline
conservative method & 0.51 & 8.49 & 6.13 \\
\hline
channel charting ($p_I = 1$) & 4.19 & 36.02 & 26.60 \\
\hline
\multicolumn{4}{|l|}{\textbf{Supervised Methods}} \\
\hline
fingerprinting ($\Delta_s = 2\, {\rm m}$) & 4.57 & 6.13 & 5.67 \\
\hline
fingerprinting ($\Delta_s = 1\, {\rm m}$) & \textbf{3.94} & 5.43 & 4.99 \\
\hline
fingerprinting ($\Delta_s = 0.5\, {\rm m}$) & 3.98 & \textbf{4.77} & \textbf{4.54} \\
\hline
\end{tabular}
\end{table}

For a quantitative evaluation of the performance, the \ac{cdf} and \ac{mae} for all considered methods are shown in Fig.~\ref{fig:cdf} and Table~\ref{tab:MAE}, respectively. It is seen that the model-based method and the unified method (with $p_I =1$) achieve the lowest estimation error for \ac{los} users; while the fingerprinting with a small $\Delta_s$ has the lowest estimation error for \ac{nlos} users and attains the highest overall performance for all users. The unified method significantly outperforms the model-based method, and achieves an estimation accuracy comparable to that of fingerprinting with $\Delta_s = 0.5 \, {\rm m}$ (within 1 m in terms of \ac{mae}). Notably, with unsupervised learning, the unified method is able to achieve a slightly better overall performance than the fully-supervised fingerprinting with $\Delta_s = 2\, {\rm m}$.

\subsubsection{Discussion on $p_I$}

\begin{figure}[t]
\centering
\begin{minipage}[t]{0.97\linewidth}
\centering
%
%
\begin{tikzpicture}

\begin{axis}[%
width=7.2cm,
height=4cm,
at={(0in,0in)},
scale only axis,
xmin=0.5,
xmax=1,
xlabel style={font=\color{white!15!black},font=\footnotesize},
xlabel={$p_I$},
ymin=0,
ymax=13,
yminorticks=false,
ylabel style={font=\color{white!15!black},font=\footnotesize},
ylabel={MAE (m)},
axis background/.style={fill=white},
xmajorgrids,
ymajorgrids,
xticklabel style={font=\footnotesize},
yticklabel style={font=\footnotesize},
legend style={fill opacity=0.5, draw opacity=1, text opacity=1, at={(1,1)},anchor=north east,legend cell align=left, align=left, font=\scriptsize, legend columns=2, draw=white!15!black}
]

\addplot [color=blue, dotted, line width=1pt]
  table[row sep=crcr]{%
0.5	    0.5099383\\
0.55	0.5099383\\
0.6 	0.5099383\\
0.65	0.5099383\\
0.7 	0.5099383\\
0.75	0.5099383\\
0.8 	0.5099383\\
0.85	0.5099383\\
0.9 	0.5099383\\
0.95	0.5099383\\
1   	0.5099383\\
};
\addlegendentry{conservative (\ac{los} users)}

\addplot [color=red, dotted, line width=1pt, mark size=0.8 pt, mark=o, mark options={solid, red}]
  table[row sep=crcr]{%
0.5	    6.459314\\
0.55	5.7556634\\
0.6 	3.740983\\
0.65	3.9332778\\
0.7 	2.7383718\\
0.75	2.0384927\\
0.8 	1.8860437\\
0.85	1.2917101\\
0.9 	1.215889\\
0.95	0.7595727\\
1   	0.34051138\\
};
\addlegendentry{unified (\ac{los} users)}

\addplot [color=blue, dashed, line width=1pt]
  table[row sep=crcr]{%
0.5	    8.490908 \\
0.55	8.490908 \\
0.6 	8.490908 \\
0.65	8.490908 \\
0.7 	8.490908 \\
0.75	8.490908 \\
0.8 	8.490908 \\
0.85	8.490908 \\
0.9 	8.490908 \\
0.95	8.490908 \\
1   	8.490908 \\
};
\addlegendentry{conservative (\ac{nlos} users)}

\addplot [color=red, dashed, line width=1pt, mark size=0.8 pt, mark=o, mark options={solid, red}]
  table[row sep=crcr]{%
0.5	    8.834487\\
0.55	9.208421\\
0.6 	8.771729\\
0.65	8.579371\\
0.7 	8.5753765\\
0.75	8.532417\\
0.8 	8.047796\\
0.85	8.049724\\
0.9 	7.793758\\
0.95	7.4704623\\
1   	7.668736\\
};
\addlegendentry{unified (\ac{nlos} users)}

\addplot [color=blue, line width=1pt]
  table[row sep=crcr]{%
0.5	    6.1295457\\
0.55	6.1295457\\
0.6 	6.1295457\\
0.65	6.1295457\\
0.7 	6.1295457\\
0.75	6.1295457\\
0.8 	6.1295457\\
0.85	6.1295457\\
0.9 	6.1295457\\
0.95	6.1295457\\
1   	6.1295457\\
};
\addlegendentry{conservative (all users)}

\addplot [color=red, line width=1pt, mark size=0.8 pt, mark=o, mark options={solid, red}]
  table[row sep=crcr]{%
0.5	    8.131735\\
0.55	8.18684\\
0.6 	7.2832627\\
0.65	7.2047133\\
0.7 	6.8483586\\
0.75	6.6110334\\
0.8 	6.2246933\\
0.85	6.0502024\\
0.9 	5.8475375\\
0.95	5.4848843\\
1   	5.5005045\\
};
\addlegendentry{unified (all users)}

\end{axis}
\end{tikzpicture}%
    \vspace{-1 cm}
\end{minipage}
\caption{\ac{mae} for the unified localization method with different $p_I$ (testing dataset).}
\label{fig:p_I}
\end{figure}
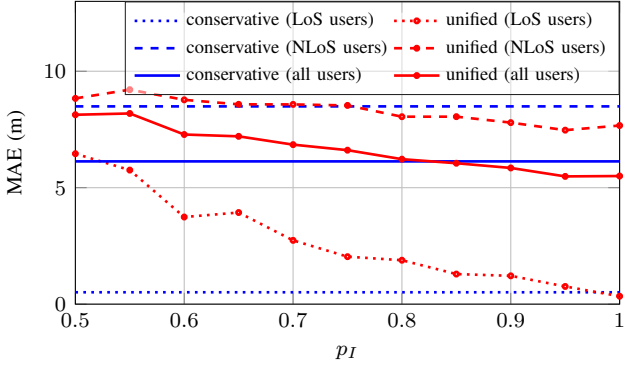

In order to explore the impact of the \ac{los}/\ac{nlos} identification accuracy, Fig.~\ref{fig:p_I} shows the \ac{mae} for the unified method with different $p_I$ and its conservative variant. It is clear (also intuitive) that the performance of the unified method relies on the identification accuracy $p_I$, and its estimation error would increase with the reduction of $p_I$. Moreover, the performance for \ac{los} users degrades much faster, while the \ac{nlos} users are more robust to the identification error. This is because the wrong identification for some \ac{nlos} users (identified as \ac{los} users by $I(\mathbf{H}_i)$ but with $f_{\rm pe}\left[ f_{\rm ce}(\mathbf{H}_i) \right] \in \mathcal{R}_{\rm NLoS}$) can be corrected based on the map information, as shown in~\eqref{eq:LoS_identification}.

It also can be seen that for a reasonable $p_I \in [0.93, 1]$ (from the existing identification methods in the literature), the unified method performs better than its conservative variant overall. There is a critical $p_I \approx 0.83$, corresponding to the intersection point of the overall \ac{mae}, below which the conservative method is a better choice. This implies that a dynamic switch between the unified and conservative methods based on $p_I$ can be adopted: when $p_I$ is larger to some threshold, e.g., 0.83, the unified method is used; otherwise, the conservative method is used.

\section{Conclusions}
In this paper, we proposed a unified localization method, along with a conservative variant, that reaps the benefits of model-based methods for \ac{los} users and \ac{ml}-based methods for \ac{nlos} users. In this method, \ac{ot} was adopted to self-generate labels such that \ac{nn} can be trained in a fully unsupervised manner without the prohibitive cost of data measurement and collection. Numerical results showed that the proposed method delivers competitive overall performance with the fully-supervised fingerprinting. Moreover, a threshold w.r.t. the \ac{los}/\ac{nlos} identification accuracy was discovered such that an appropriate choice between the unified method and its conservative variant can be made for better performance.

\bibliographystyle{IEEEtran}
\bibliography{main}

\end{document}